# Multi-Frequency Study of Spectral Indices of BL Lacertae Objects and Flat Spectrum Radio Quasars

E. U. Iyida[*], F. C. Odo, A. E. Chukwude and A. A. Ubachukwu

Astronomy and Astrophysics Research Lab., Department of Physics and Astronomy, University of Nigeria, Nsukka, Department of Physics and Astronomy, Faculty of Physical sciences, University of Nigeria, Nsukka, Nigeria
*email: evaristus.iyida@unn.edu.ng

## Abstract

We present statistical analyses of large homogenous data sample of *Fermi*-detected blazars thoroughly studied in order to reassess the relationship between flat spectrum radio quasars (FSRQs) and subclasses of BL Lacertae objects (BL Lacs) blazar populations. We discovered from the average values of γ-ray ($\alpha_{\gamma-ray}$) and X-ray ($\alpha_{X-ray}$) spectral indices that the sequence of distribution is indicative of blazar orientation scheme. Analyses of FSRQs and BL Lacs data show difference in the shape of γ-ray and X-ray indices: significant anti-correlation ($r\sim > -0.79$) exists between $\alpha_{\gamma-ray}$ and $\alpha_{X-ray}$ spectral indices. The spectral energy distributions of the blazar subclasses show that FSRQs and BL Lacs have similar spectral properties which can be unified through evolutionary sequence. Nevertheless, there is a significant difference between the shapes of X-ray and γ-ray spectra of blazars suggestive that different mechanisms are responsible for spectral variations in the two energy bands. All these results suggest that there is a form of a unified scheme for all blazars.

## 1.0 Introduction

Blazars are about 10% of active galactic nuclei (AGNs) and the most extreme subsets. Their broadband emissions from low energy radio to high energy γ-rays are dominated by non-thermal emissions produced by the relativistic plasma jets aligned to the line of sight (Blandford and Rees, 1978). Blazars comprise BL Lacertae objects (BL Lacs) and flat spectrum radio quasars (FSRQs) based on their spectroscopic properties. FSRQs have strong emission lines, while BL Lacs have very weak or no emission lines at all (Fan, 2005; Padovani, 2007). In addition, while all FSRQs are low frequency peaking sources, BL Lacs are divided into high-synchrotron (HSPs), intermediate synchrotron (ISPs) and low synchrotron (LSPs) peaked sources, based on their synchrotron peak frequencies with the values in the range: $\log \nu^{syn}_{peak} > 15$ (Hz), $14 \leq \log \nu^{syn}_{peak} \leq 15$ (Hz) and $\log \nu^{syn}_{peak} < 14$ (Hz) respectively (see, e.g. Ackermann *et al.* 2011; 2015; Ajello et al. 2020). Measuring and studying the emission spectra of these sources is essential in determining their compositions as well as relationship among them. The observed spectra which are measured through a single filter are normally reduced to the source's rest frame called intrinsic or emitted spectra. The spectral energy distributions (SEDs) of blazars give two peaks in a frequency – luminosity ($v$ - $vf$) representation, often attributed to synchrotron process at low (radio to X-ray) frequencies and inverse Compton process at high (X-ray to $\gamma$-ray) frequencies (see, e.g. Padovani *et al.* 2001; Finke, 2013). Due to increased observational and better theoretical understanding of the nature of these sources, a unification scheme has been developed that fundamentally attempts to explain the relationship among the blazar subclasses. The study of this relationship has been an unresolved problem in blazar research in a couple of decade and is still under discussion (see, e.g., Ghisellini et al., 1998; Fossati et al., 1998; Abdo et al. 2009; 2010c; Palladino et al., 2019; Iyida et al. 2019; 2020; Ouyang et al. 2021). The success of the popular relativistic beaming paradigm rests on its ability to satisfactorily explain the chameleon nature (especially, rapid variability of radiation fluxes over short and long timescales) of the SEDs of blazars. This includes flat spectrum radio quasars at high luminosity regime and BL Lacertae Objects at low luminosity end. Thus, it appears to give explanation to the observed differences between BL Lacs and FSRQs (e.g. Odo et al. 2012), as well as other related sources (Ubachukwu and Chukwude, 2002). The long-standing orientation-based unification scheme proposes that BL Lacs and FSRQs are counterparts to Fanaroff-Riley type I and II (FR I and FR II) radio galaxies, respectively, based on similar spectra, morphologies and range in extended radio luminosity (Urry and Padovani1995). The amount of relativistic beaming and intrinsic power are low in BL Lacs compared to FSRQs (e.g. Ghisellini et al., 1993; Chen et al., 2016; Odo and Aroh, 2020) thus, implying some intrinsic differences. The relativistic jet angle that is aligned to the line of sight may pointedly give rise to differences in the properties of diverse classes of blazars due to increase in relativistic beaming effect (see, e.g., Vagnetti et al., 1991; Finke, 2013). Thus, the hypothesis that BL Lac objects represent FSRQs that have highly boosted continuum is gradually doubted knowing that the extent of intrinsic power and relativistic beaming are higher in more of FSRQs than in BL Lac objects (e.g. Padovani 1992; Ghisellini et al., 2017; Odo et al. 2017), thus, indicating that there are inherent differences between them

On the contrary, it is generally believed that BL Lacs and FSRQs are objects in different forms with the same physical process but differ only in bolometric luminosity (e.g. Ghisellini et al., 1998; Fossati et al., 1998). This has led to the popular blazar sequence that appears to have dominated discussions on unification of blazars in recent investigations (e.g. Fossati et al., 1998; Meyer et al., 2011; Finke, 2013; Mao et al., 2016; Iyida et al. 2020; 2021) indicating that the differences between the blazar subclasses are due to the differences in bolometric luminosity. Meanwhile, according to the broadband SEDs, the orientation and unification scheme of blazars suggest a sequence in which the frequency at the peak of synchrotron emission is anti-correlated with the synchrotron peak luminosity ($L_{syn}$). In the same vein, a significant side of the blazar sequence that has attracted some authors' attention (e.g. Giommi et al., 2012; Finke, 2013; Odo et al., 2017; Nalewajko and Gupta, 2017; Odo and Aroh, 2020) is the relationship between the high and low-energy components of the SEDs. Fossati et al. (1998) introduced a broadband parameter, called, γ-ray dominance ($D_g$), defined as the ratio of γ-ray luminosity ($L_g$) to the synchrotron peak luminosity ($L_{pk}$) and found strong anti-correlation between the parameter and synchrotron peak frequency for the EGRET blazars, which the authors used to argue for the blazar sequence (see, also Odo and Aroh, 2020). Another parameter that relates the low to high energy components of the SED is the spectral index (α). This parameter is often defined between two frequencies in the low and high energy regimes (e.g. Abdo et al. 2010a). A major argument against this parameter is that it is somewhat redshift-dependent (e.g. Athreya and Kapahi, 1998), which needs to be corrected for in source samples. Therefore, assuming BL Lacs and FSRQs differ in orientation, the difference in their SEDs can also be accommodated in the beaming models. In this sense, blazar sequence posits that misaligned blazars will drop in spectral luminosity according to the decrease in Doppler boosting with increasing viewing angle and synchrotron peak frequency (Meyer et al., 2011). However, it was thought that BL Lacs are seen through intervening galaxies so that microlensing effects lead to higher boosting than in FSRQs (Ostriker and Vietri, 1985). The fact that luminosity of FSRQs in the radio band is systematically higher than that of BL Lacs seems to favour a unified scheme via orientation, but somewhat contradictory to the initial supposition that BL Lacs represent FSRQs with the most highly boosted continuum (e.g. Vagnetti et al., 1991). In spite of many explorations with GHz band, the composite spectral properties in terms of the blazar sequence at multi-frequency level is less explored. Therefore, knowing that the quantifiable knowledge of the spectral properties is a crucial orientation parameter that will not only provide qualitative information but also give consistency tests of the orientation based unification scheme for the radio loud active galactic nuclei, a good assessment of the blazar continuity is to explore its validity in the high and low-energy peaked systems. In this paper, we investigate the spectral indices of a significant sample of blazars from the low to high energy frequencies using the *Fermi*-Large Area Telescope (*Fermi*-LAT) data. This is in order to test and ascertain the validity of the blazar sequence in these energy bands. Section 2 discusses basic theoretical concepts while section 3 describes the data sample and the results of the relationship between X-ray spectral index ($\alpha_{X-ray}$) and γ-ray spectral index ($\alpha_{\gamma-ray}$) and other correlations in different wave bands. In Section 4, we gave the discussion. The main conclusion is summarized in Section 5. Throughout this paper, we adopted the standard cold dark matter $(\Lambda - CDM)$ cosmology with Hubble constant with $H_0$ = 71 km s$^{-1}$ Mpc$^{-1}$, $\Omega_M$ = 0.27 and $\Omega_\Lambda$ = 0.73. All relevant data are adjusted based on this concordance cosmology.

**2.0 Basic Theoretical Concepts**

The long wavelength emissions from radio to X-ray are mostly dominated by the synchrotron mechanism coming from the outer parsec scale region of the jet while the short wavelength emissions from X-ray to γ-ray are linked to inverse Compton scattering process coming from the inner parsec scale region of the jet (see, e.g. Finke, 2013). Several authors adopted some proxy parameters, which depend on broadband spectral energy distribution, as well as observing frequency to study the blazar sequence (Fossati et al. 1998; Finke, 2013; Nalewajko and Gupta, 2017). Among the proxy parameters that are frequently used is the effective broadband spectral index ($\alpha_{1-2}$), usually defined between two frequencies (e.g. Leddden and Odell 1985) as:

$$\alpha_{1-2} = -\frac{\log\left(\frac{S_1}{S_2}\right)}{\log\left(\frac{v_1}{v_2}\right)}, \tag{1}$$

where $S_1$ and $S_2$ are the observed monochromatic fluxes at frequencies $v_1$ and $v_2$ respectively. Previous studies of the blazar sample suggest a relationship between these multi-frequency emissions and relativistic beaming (Cellone *et al.*, 2007; Savolainen et al., 2010; Finke 2013). Odo *et al.*, (2017) and lately, Iyida et al., (2021) provide evidence that orientation and relativistic beaming are imperative in explaining variations of high and low energy emissions from blazars. This is evidently in agreement with Giommi *et al.* (2013) that there is a relationship between high and low energy flux of blazar samples. However, the fact that high energy sources have small linear sizes implies that high energy emissions are from the core and are relativistically beamed (Savolainen *et al.* 2010).

The generally accepted model of radio sources indicates that the intrinsic properties of these radio sources are modified by Doppler beaming due to relativistic speed and orientation effect. This relativistic beaming model has been remarkably effective in explaining the rapid variability of monochromatic flux from blazars. It is observed that BL Lacs and FSRQs are strongly beamed, so, the relationship between them is complicated. Thus, it becomes necessary to consider their intrinsic properties based on the Doppler corrected data. The observed spectral flux ($S_{obs}$) is related to the intrinsically emitted flux ($S_{in}$) (see, e.g. Lind and Blandford 1985; Pei *et al.* 2019) by $S_{obs} \approx \delta^{n+\alpha} S_{in}$, where $n$ is a jet model dependent factor which is either 2 or 3 for continuous jet or blob models, respectively, and $\delta$ is Doppler factor, This observed spectra flux must be Doppler and $k$-corrected to the intrinsic spectra flux using

$$K = S_{obs}(1+z)^{\alpha-1}\delta^{-3-\alpha(v)} \tag{2}$$

with $\alpha$ being the spectral index, while $z$ is the redshift of the source. However, knowing that the bulk Lorentz factor of the plasma increases with distance from the centre as the radiation is beamed along the jet, the Doppler factor can be estimated for different objects (see, e.g. Xie et al. 1991; 2001). Thus, this implies that the composite spectral indices obtained from the observed data cannot represent the fundamental property. Thus, we look at the spectral properties of these objects from the low-energy component dominated by synchrotron process to the high-energy component dominated by inverse Compton process by calculating the composite spectral indices using

$$\alpha_{1-2} = -\frac{\log\left(\frac{S_1}{S_2} \cdot K\right)}{\log\left(\frac{\nu_1}{\nu_2}\right)} \qquad (3)$$

where $K$ is the total $k$-correction factor given by equation (2). This composite spectral indices give information about the relationship between the blazar subclasses.

**3.0 Data Sample and Results**

**3.1 Sample**

The data sample employed in the current analysis is based on third *Fermi* Large Area Telescope (*Fermi*-LAT) AGNs catalogue (3LAC) as compiled by Ackermann et al. (2015) and Acero et al. (2015). From the catalogue, we selected a sample of 1081 blazars with clear optical identification, namely, 461 FSRQs and 680 BL Lacs. First, from literature, we collected for the current sample, the multi-frequency spectral indices in the radio ($\alpha_R$), optical ($\alpha_O$), X-ray ($\alpha_X$) and γ-ray ($\alpha_\gamma$) bands. In particular, we cross-matched the sources in our sample with those in the second *Fermi*-LAT (2LAC) catalogue (Nolan et al. 2012), for information on X-ray and γ-ray photon spectral indices ($\Gamma$) from where their spectral indices were obtained. The observed monochromatic luminosities ($L_m$) at radio 1.4 GHz, optical $R$ band ($4.68 \times 10^{14}$ Hz), X-ray at 1 keV and 1.0 GeV γ-ray data were obtained from Acero et al. (2015) and we calculated the spectral flux density using $S = \frac{L_m}{4\pi d_L^2}$ where $d_L$ is the luminosity distance expressed following (Alhassan et al. 2019) as $d_L = H_o^{-1} \int_0^z \left[(1+z)^2(1+\Omega_m z) - z(2+z)\Omega_\Lambda\right]^{-\frac{1}{2}} dz$. As expected, several objects in the 3LAC were missed out in the selection due to their absence in the 2LAC. Similarly, some sources listed in 2LAC are also missing in the 3LAC, perhaps due to variability of radiation flux during the duty cycles presented in the two catalogues. Nevertheless, a good number of sources that were unassociated in the 2LAC are now associated with blazars in the 3LAC. More so, some 2LAC sources have changed classifications in 3LAC. These are attributable to improved data. The effect of all these is that our sample is significantly less than the 3LAC sample presented by Ackerman et al. (2015) and Acero et al. (2015). The final sample consists of 401 blazars which include 253 BL Lacs and 148 FSRQs, with complete information on the relevant parameters. For all the sources, the composite spectral indices:

Optical-X-ray ($\alpha_{OX}$), Radio-X-ray ($\alpha_{RX}$), Radio-Optical ($\alpha_{RO}$) and X-ray-γ-ray ($\alpha_{X\gamma}$) were computed. The Doppler factors of these objects were estimated in different wavebands from literature (Xie et al., 1991; 2001; Ghisellini et al., 1993; Dondi and Ghisellini, 1995; Cheng et al., 1999). For a source without available redshift and Doppler factor, the average value of the corresponding group is used for the calculation.

**3.2 Distributions of γ-ray and X-ray Spectral Indices**

We show in Figure 1 the distributions of the sample in both $\alpha_{\gamma-ray}$ and $\alpha_{X-ray}$ spectral index. It is however clear from Fig. 1(a) that FSRQs, on average, have the tendency to possess higher values of $\alpha_{\gamma-ray}$ than the BL Lacs subclasses. The average values are 1.38, 1.16, 1.12 and 0.92 for FSRQs, LSPs, ISPs and HSPs, respectively. Hence, the average values of the $\alpha_{\gamma-ray}$ for the blazar subclasses follow the relation: $\langle\alpha_{\gamma-ray}\rangle|_{HSPs} < \langle\alpha_{\gamma-ray}\rangle|_{ISPs} < \langle\alpha_{\gamma-ray}\rangle|_{LSPs} < \langle\alpha_{\gamma-ray}\rangle|_{FSRQs}$ which is suggestive of spectral sequence of blazar unified scheme. However, FSRQs and BL Lacs are respectively distributed in unimodal and multimodal configurations respectively. Nevertheless, the multimodal distributions of BL Lacs arise from the fact that there are different subclasses of BL Lacs. This result can be interpreted to mean that in terms of γ-ray spectra index, different subclasses of blazars may exist.

In the case of $\alpha_{X-ray}$ shown in Fig. 1b, FSRQs range from 0.15 to 1.47 peaking at 0.57 with average value of 0.72. However, BL Lacs range from 0.64 to 2.18 peaking at different values for LSP, ISP and HSP objects with average values of 1.31, 1.24, and 1.72 for LSPs, ISPs and HSPs, respectively. We find that the average values of the $\alpha_{X-ray}$ for the blazar subclasses follow the relation:, $\langle\alpha_{X-ray}\rangle|_{HSPs} > \langle\alpha_{X-ray}\rangle|_{ISPs} > \langle\alpha_{X-ray}\rangle|_{X-ray} > \langle\alpha_{X-ray}\rangle|_{FSRQs}$, a trend that is more or less a mirror image of the $\alpha_{\gamma-ray}$ distribution, and also suggests a sequence for the blazar sample. Further statistical test reveals that FSRQs and BL Lac subclasses are not nicely fitted to log-normal distribution with skewness (*μ*) between - 0.03 and 0.02. The mirrored spectral shapes in X-ray and γ-ray band suggests that different mechanisms are responsible for the variations in γ-ray and X-ray spectral indices of all subclasses of blazars. Perhaps, it can be argued that in the blazars, X-ray emission arises from synchrotron mechanism, while the γ-ray emission is due to inverse Compton process (e.g. Iyida et al. 2020).

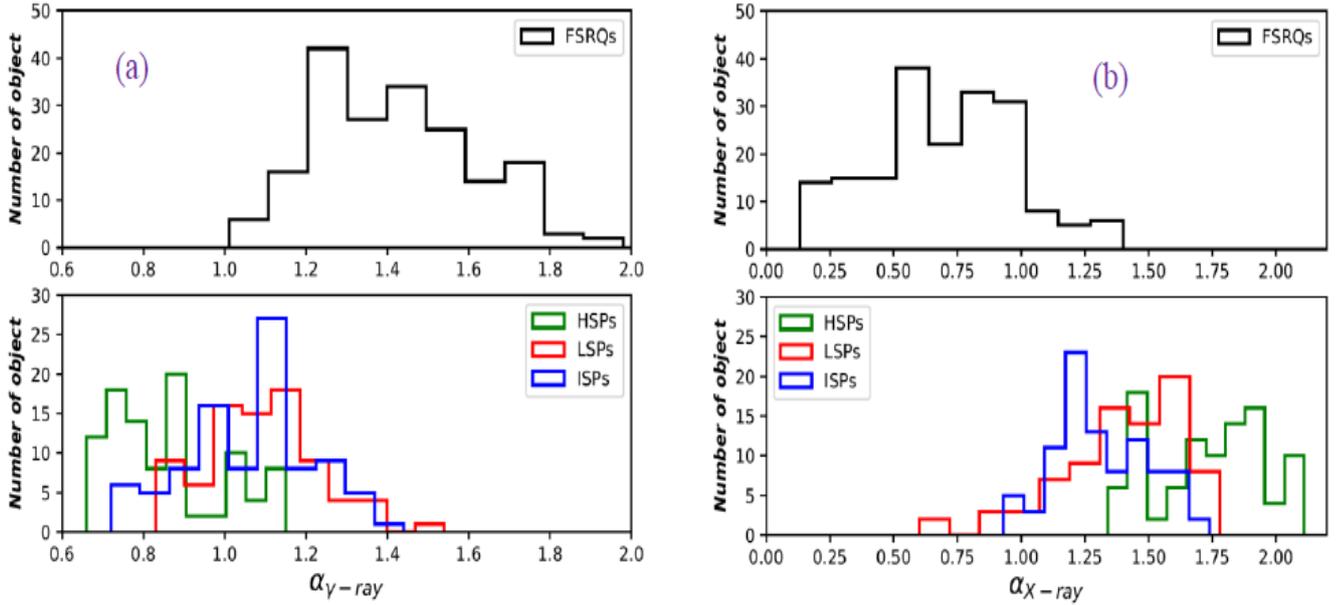

**Figure 1:** Histogram showing the distributions of (a) $\alpha_{\gamma-ray}$ (b) $\alpha_{X-ray}$ of blazars in our sample

### 3.3 Correlations among Spectral Parameters

The spectral parameters and composite spectral indices of our sample were investigated using the Pearson correlation theory. This is aimed at understanding the processes responsible for the γ-ray and X-ray emissions in blazars. For two variable data sets, $x_i$ and $y_i$, we use the linear regression fitting to analyse their correlations, expressed as $y = (k \pm \Delta k)x + (k_0 \pm \Delta k_0)$, where $k$ is the slope and $k_0$ is the intercept with their errors stated. The Pearson's correlation coefficient $r$ is expressed (see, Press 1994; Pavlidou et al. 2012) as:

$$r = \frac{\sum(x_i - \bar{x})(y_i - \bar{y})}{\sqrt{\sum(x_i - \bar{x})}(\sqrt{\sum y_i - \bar{y}})} \quad (4)$$

where $\bar{x}$ and $\bar{y}$ are the mean values of $x_i$ and $y_i$ with $(x_i, y_i)$ corresponding to $(\alpha_{\gamma-ray}, \alpha_{X-ray})$ respectively. The level of the significance (z) is determined by using a quantity (z) defined as $z = \frac{b}{\sqrt{[var(b)}}$ where *Var(b)* is variance and *b* the slope of regression line. To test for the correlations between the spectral parameters of our sample, the $\alpha_{\gamma-ray} - \alpha_{X-ray}$ plot is shown in Figure 2. FSRQs are exclusively located at high $\alpha_{\gamma-ray}$, low $\alpha_{X-ray}$ end while BL Lacs span a wide range, with ISPs and LSPs objects overlapping with FSRQs at intermediate values. The distribution of the objects on the $\alpha_{\gamma-ray} - \alpha_{X-ray}$ plane is consistent with the proposed blazar sequence. While most BL Lacs have flat spectra $|\alpha_{\gamma-ray}| < 0.75$, FSRQs show steep spectra $|\alpha_{\gamma-ray}| > 0.75$ with a wider range. This is an indication of the range of $\gamma-ray$ emission mechanisms in FSRQs. It can be argued that the trend

observed from FSRQs up to HSPs follows a system which is indicative of the orientation unified scheme and that related mechanisms control these objects.

Furthermore, it is known that considerable uncertainties characterize the statistical analyses of large data sample of extragalactic sources. Thus, in view of the extent of uncertainties in the γ-ray and X-ray spectral indices of blazars, we posit that a more objective investigation of their relationship might require the use of the average values of the parameters obtained from carefully chosen bins. The bining was done over the X-ray spectral indices as follows: $\alpha_{X\text{-ray}} \leq 0.25$; $0.25 < \alpha_{X\text{-ray}} \leq 0.50$; $0.50 < \alpha_{X\text{-ray}} \leq 0.75$; $0.75 < \alpha_{X\text{-ray}} \leq 1.00$; $1.00 < \alpha_{X\text{-ray}} \leq 1.25$; $1.25 < \alpha_{X\text{-ray}} \leq 1.50$; $1.50 < \alpha_{X\text{-ray}} \leq 1.75$; $1.75 < \alpha_{X\text{-ray}} \leq 2.0$ and $\alpha_{X\text{-ray}} > 2.0$. The average values of γ-ray and X-ray spectral indices were calculated for each bin. The standard errors of the average values of these parameters were also calculated. The plot of the average values of γ-ray spectral indices against the average values of X-ray spectral indices was superimposed on the γ-ray and X-ray spectral indices as shown in Figure 2. Linear regression analysis of $\alpha_{\gamma\text{-ray}}$ and $\alpha_{X\text{-ray}}$ data gives $\alpha_{\gamma-ray} = -(0.64 \pm 0.20)\alpha_{X-ray} + (1.83 \pm 0.02)$ with a correlation coefficient ($r \sim > 0.79$). The correlation is found to be statistically significant at 98% confidence level and suggests that similar effects are responsible for variations in the parameters for both BL Lacs and FSRQs. This depends on a factor that changes progressively from HSPs to FSRQs through LSPs and ISPs.

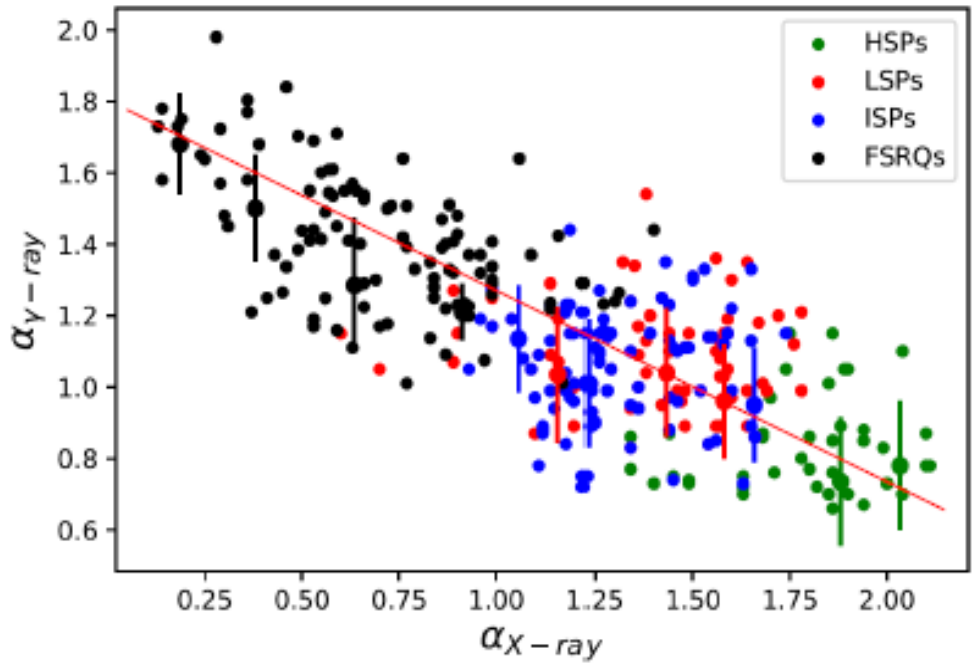

**Figure 2: Plot of $\alpha_{\gamma-ray}$ against $\alpha_{X-ray}$ for blazar subsets in our sample**

## 3.4. Correlations between composite Spectral Indices

Here, the relationship that exists among the composite spectral indices of our whole data sample is investigated. The average values of composite spectral indices were calculated with the standard errors of the mean values. The scatter plots are superimposed as shown in Fig. 3 (a – f). However, the slope $k$ of the plots, intercept $k_0$, correlation coefficient $r$ and chance probability $p$ and their errors are all listed in Table 1. We argue that the significant correlations obtained implies that common factor that change linearly between the subclasses is responsible for the variation and that these blazar subclasses can be unified through their broadband spectral indices.

Table 1: Results of Linear regression analyses of composite spectral indices for the whole sample

| Parameter | $k$ | $\Delta k$ | $k_0$ | $\Delta k_0$ | $r$ | $p$ |
|---|---|---|---|---|---|---|
| $\alpha_{XY} - \alpha_{RX}$ | -1.13 | 0.20 | 1.21 | 0.02 | -0.76 | $< 10^{-4}$ |
| $\alpha_{XY} - \alpha_{RO}$ | -1.16 | 0.30 | 1.24 | 0.03 | -0.74 | $< 10^{-4}$ |
| $\alpha_{OX} - \alpha_{RO}$ | -1.60 | 0.30 | 2.12 | 0.02 | -0.43 | $< 10^{-4}$ |
| $\alpha_{RX} - \alpha_{RO}$ | 0.75 | 0.20 | 0.52 | 0.02 | 0.58 | $< 10^{-4}$ |
| $\alpha_{RO} - \alpha_{\gamma-ray}$ | 1.32 | 0.20 | 0.58 | 0.02 | 0.67 | $< 10^{-4}$ |
| $\alpha_{RX} - \alpha_{X-ray}$ | -3.40 | 0.20 | 2.30 | 0.02 | -0.55 | $< 10^{-4}$ |

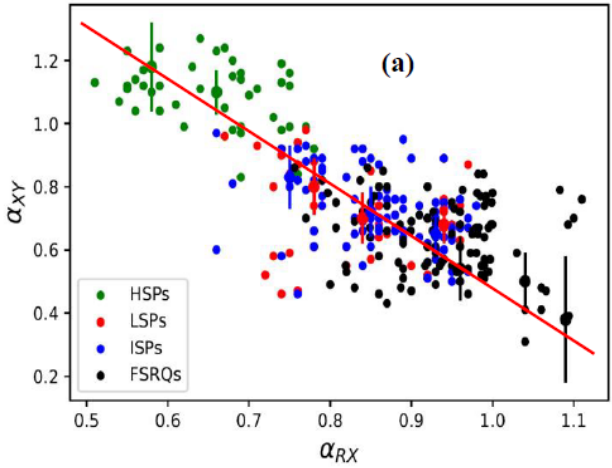
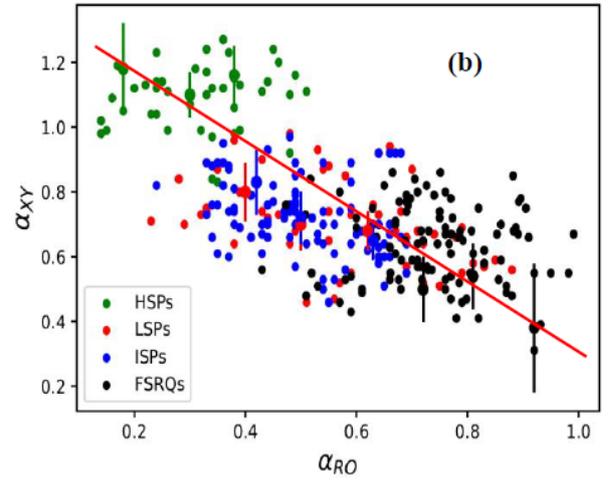

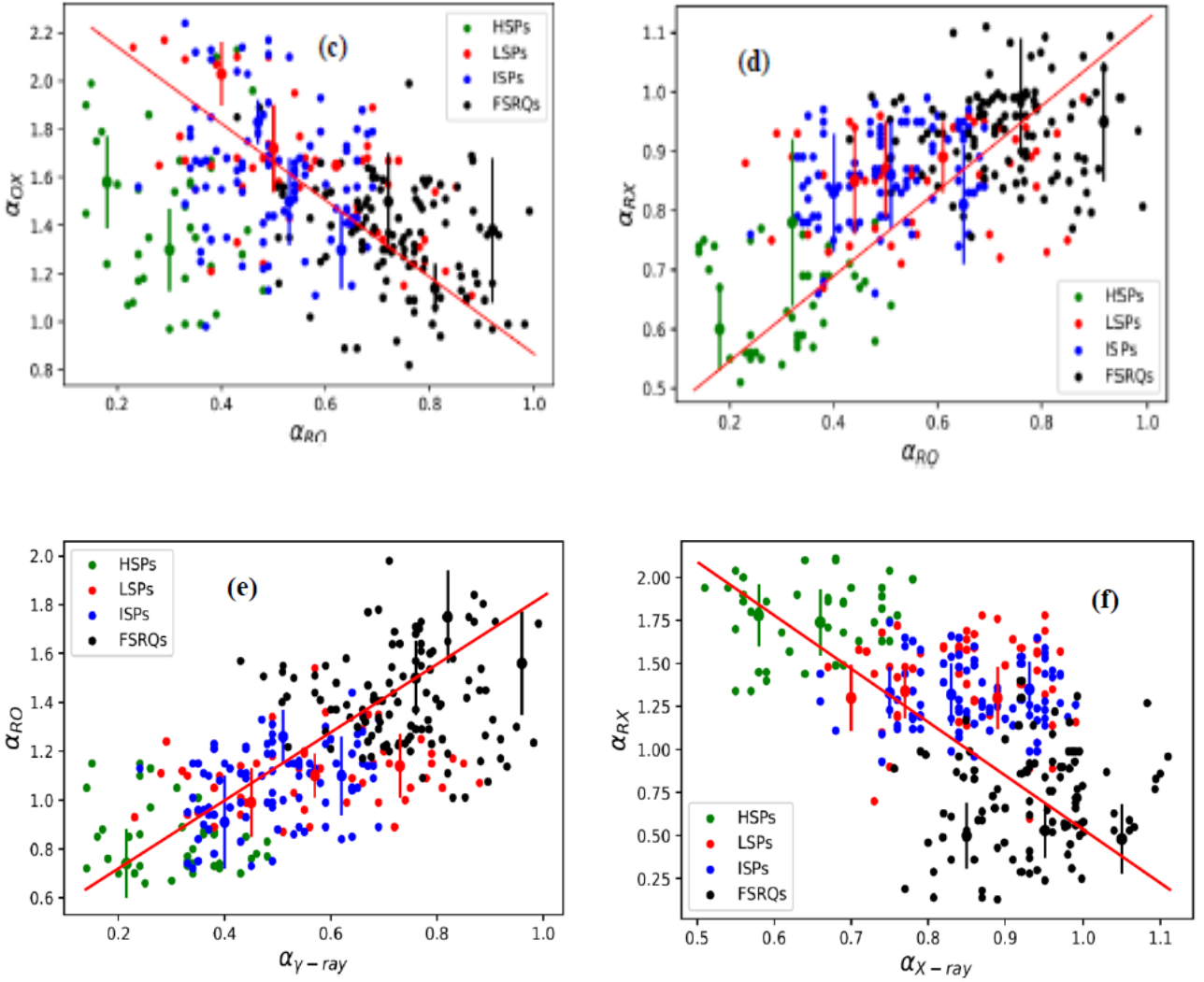

Figure 3: Plots of (a) $\alpha_{X\gamma}$ against $\alpha_{RX}$ (b) $\alpha_{X\gamma}$ against $\alpha_{RO}$ (c) $\alpha_{OX}$ against $\alpha_{RO}$ (d) $\alpha_{RX}$ against $\alpha_{RO}$ (e) $\alpha_{RO}$ against $\alpha_{\gamma-ray}$ (f) $\alpha_{RX}$ against $\alpha_{X-ray}$ for the blazar samples

**3.5 Correlations between γ-ray and X-ray Spectral Indices with the redshift**

A major problem against the use of spectral index in unification studies is that the parameter is somewhat redshift dependent (Athreya and Kapahi,1998), with the spectrum being steeper at high redshifts than at low redshifts, which would introduce some ambiguities in the analyses. Thus, we examined the implications of this redshift effect of the present sample in a considerable detail and the uncertainties in the values equally taken care of. The plots of α$_{γ-ray}$ and α$_{X-ray}$ against redshift with the superimposed plot of their mean values are shown in Figs. 4a and 4b while the regression results are given in Table 2. It is apparent from the figures that while a vast majority of FSRQs are located at high redshifts, BL Lacs are mainly located at low redshifts, suggestive of a form of selection effect in the sample. However, it is known that blazar sequence is derived from a highly (observationally) biased sample, and might not reflect a true systematic trend in the intrinsic properties of blazars. Also, we noted that the instrumental bias in the sample of blazars is not uniform since measurements are normally taken with different instruments having different sensitivities. Despite these shortcomings the results of the current analyses are encouraging, as it provides evidence that the presence of the spectral trend with redshift, could be the result of a peculiar parent population, though the instrumental bias has to be carefully studied. Also, it could be argued that the effects of γ-ray in the intergalactic medium due to extragalactic background light EBL may cause the observed γ-ray spectra to be larger at high redshifts than their intrinsic states (See, Behera and Wagner, 2009; Alonso, 2019). It could thus produce the effective spectral trend with redshift as seen in the current sample of blazars. Nevertheless, our results show smooth transition from HSPs to FSRQs through LSPs and ISPs with the sources quite overlapping on the plane. This seems to suggest that BL Lacs and FSRQs can be unified through evolution. On the other hand, while the redshift dependence of the spectral index is positive in the γ-ray band, it is negative in X-ray band. This again supports the supposition that different mechanisms are responsible for X-ray and γ-ray spectra of blazars.

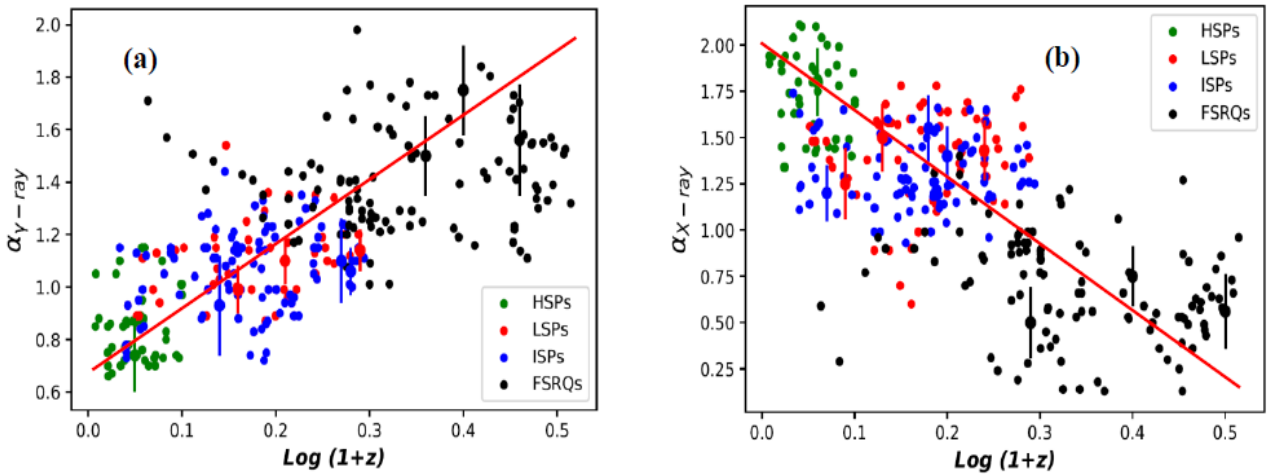

**Figure 4:** Plots of (a) $\alpha_{\gamma-ray}$ and their mean values (b) $\alpha_{X-ray}$ and their mean values against the redshift of our sample

**Table 2:** Results Linear regression analysis of $\alpha_{\gamma-ray}$ and $\alpha_{X-ray}$ with $z$ for the whole sample

| Parameter | k | Δk | k₀ | Δk₀ | r | p |
|---|---|---|---|---|---|---|
| $\alpha_{\gamma-ray} - \log(1+z)$ | 3.13 | 0.20 | 0.64 | 0.02 | 0.63 | $< 10^{-4}$ |
| $\alpha_{X-ray} - \log(1+z)$ | -1.16 | 0.30 | 2.25 | 0.03 | -0.61 | $< 10^{-4}$ |

## 4. Discussion

The unified idea about the active galactic nuclei is mainly centred around the black hole that is frequently accreting matter. The different radiations originating from blazars are generally known to be related to the relativistic jets. The spectral properties of the blazars are believed to be associated closely with electronic energy distributions and indices of the electrons (Ghisellini et al., 1998, Sambruna et al., 1996). The knowledge of the relationships among the various blazar subclasses is important since it can improve our knowledge of the basis of blazars and their evolutions. The difference in the composite spectral indices distribution between blazars subsets may be attributed to different intrinsic backgrounds within the nucleus of blazars. The blazars subsets are believed to have inherent environments with physical differences. Though BL Lacs have cleaner environments with no dust particles, FSRQs are full of both interstellar gas and dust, and thus, the accretion rates into the centre of supermassive black hole will be higher in them than in BL Lacs. The spectral properties of the blazars are assumed to come from more than one component as demonstrated by the relations and $\langle \alpha_{\gamma-ray} \rangle |_{\text{ISP}} > \langle \alpha_{\gamma-ray} \rangle |_{\text{LSP}} > \langle \alpha_{\gamma-ray} \rangle |_{\text{HSP}}$ and $\langle \alpha_{X-ray} \rangle |_{\text{ISP}} < \langle \alpha_{X-ray} \rangle |_{\text{LSP}} < \langle \alpha_{X-ray} \rangle |_{\text{HSP}}$ implying that the different physical origins of γ-ray emissions in BL Lacs and FSRQs may account for their distinct γ-ray properties (see, e.g. Fan et al., 2016; Boula et al., 2019). For BL Lacs, the high energy emissions are generally believed to originate from pure synchrotron self-Compton (SSC) process (e.g., Mastichiadis and Kirk, 1997; Zhang et al., 2012), while for FSRQs, it is usually presumed to originate from the synchrotron self-Compton and External-Compton (SSC + EC) radiation processes (e.g. Böttcher and Chiang, 2002; Ghisellini et al., 2011; Yan et al., 2014; Zheng et al., 2017). This indicates that the FSRQs jet has a complex external physical environment, which may lead to their complex physical properties (see e.g., Kang et al., 2019). Further statistical results show that LSPs and ISPs are higher in $\alpha_{\gamma-ray}$ than in $\alpha_{X-ray}$; indicating that there may be differences in the emission mechanisms within the blazar subsets. For instance, the $\alpha_{\gamma-ray}$ of FSRQs is greater than that of the of BL Lacs (see e.g., Abdo et al., 2010b; Ajello, 2020) and this may be the result of the spectrum being superposed with other spectral components for the BL Lacs.

In the $\alpha_{\gamma-ray}$ - $\alpha_{X-ray}$ plane, it is shown that FSRQs and HSPs are well divided while LSPs and ISPs are indistinguishable and bridge them. FSRQs have large values of $\alpha_{\gamma-ray}$ but less values for $\alpha_{X-ray}$. This anti-correlation is possible because, the $\alpha_{X-ray}$ of FSRQs is known to be dominated by the synchrotron self-Compton and external emissions of electrons with very low energy and this leads to insignificant $\alpha_{X-ray}$ while HSPs have steep $\alpha_{X-ray}$ that occurs for the reason that it is fully controlled by the synchrotron self-Compton emissions of highly energetic electrons. However, the locations of these blazar subsets is such that follows the blazar sequence. The neat separation between blazar subclasses in the composite spectral indices plane are really remarkable. The significant correlations in the composite spectral indices of FSRQs and BL Lacs subclasses implies that similar processes occur in all these objects within the same physical environments. Likewise, the inconsistencies among the three subclasses of BL Lacs may perhaps be attributed to intrinsically different locations within the nucleus of blazars as well as the different stages of cooling for the various subclasses. Thus, they do not have the same composite spectral indices. The close relationships shown by the composite spectral indices and the $\alpha_{\gamma-ray}$ and $\alpha_{X-ray}$ is notable since the FSRQs are known to be associated with FR II with complex physical environments originating from both synchrotron self-Compton and external radiations while the BL Lacs subclasses are normally found in the FRIs with emission coming from mainly synchrotron self-Compton radiations. The Continual trend going from FSRQs to HSPs which is consistent with blazar sequence indicates that related physical process occur in all these objects. There is a tendency for the redshifts to become smaller with increase in $\alpha_{X-ray}$ of the blazars from FSRQs to HSPs. This is rather different from that of $\alpha_{\gamma-ray}$ which has the redshifts increase in proportion from HSPs to FSRQs. Though, $\alpha_{X-ray}$ of FSRQs and BL Lacs are negatively associated with redshifts, it is rather positive for $\alpha_{\gamma-ray}$. Our results show that the $\alpha_{\gamma-ray}$ and $\alpha_{X-ray}$ peak at high redshifts for FSRQs but the reverse is the case for BL Lacs. The large scatter in $\alpha_{X-ray}$ and $\alpha_{\gamma-ray}$ at high redshifts suggests that luminosity selection effect plays important role (e.g. Alhassan et al., 2013; Odo et al., 2014). Actually, the effect of the luminosity redshift is very significant such that it can swamp any effect that comes from relativistic beaming of FSRQs at high redshifts. Thus, the sequence of variation of the $\alpha_{X-ray}$ and $\alpha_{\gamma-ray}$ may be an artefact of the luminosity selection effect at high redshifts. The presence of a fraction of FSRQs that have the same values with LSPs and ISPs does not change the trend from HSP to FSRQs rather they are located in such a manner that is consistent with a unified scheme between BL Lacs and FSRQs through evolution.

## 5. Conclusion

The results of the spectral properties of some *Fermi* selected blazars are presented. We obtained strong anti-correlation between the $\alpha_{\gamma-ray}$ and $\alpha_{X-ray}$ for our sample and thus, it supports the unified scheme for blazars. We also found that HSPs and FSRQs occupy different regions in the color-color diagrams while ISPs, LSPs are mixed together and links them. This plausibly suggests that ISPs, LSPs and FSRQs have intrinsic and comparable spectral properties while HSPs have distinct properties. The tighter $\alpha_{X-ray} - z$ and $\alpha_{\gamma-ray} - z$ connection for BL Lacs, combined with the flatter slope of redshift of FSRQs suggest that the core emissions for BL Lacs is more dominant at higher redshifts than that for FSRQs.

**Acknowledgement**

We sincerely thank an anonymous referee for constructive, helpful comments and suggestions which greatly improved the paper. This work was done with the aid of financial support from the National Economic Empowerment Strategy (NEEDS) of the Federal Government of Nigeria (FGN).

**REFERENCES**

Abdo A A, Ackermann M, Ajello M, Atwood WB, *et al.* Bright Active Galactic Nuclei Source List from the First Three Months of the Fermi Large Area Telescope All-Sky Survey. Astrophysical J., 2009; 700: 597-62265

Abdo A A, Ackermann M, Agudo I, Ajello M, *et al*. The Spectral Energy Distribution of Fermi Bright Blazars. Astrophysical J., 2010c; 716, 30-70

Abdo A A, Ackermann M, Ajello M, Atwood WB. et al. Spectral Properties of Bright Fermi-Detected Blazars in the Gamma-ray Band. Astrophysical J., 2010a; 710, 1271-1285

Abdo AA, Ackermann M, Ajello, M, Allafort A. et al. The First Catalog of Active Galactic Nuclei Detected by The Fermi Large Area Telescope. Astrophysical J., 2010b; 715, 429–457

Acero F, Ackermann M, Ajello M, Albert A, Atwood W B. Fermi large area telescope third source catalogue. ApJS. 2015; 218: 23-64

Ackermann M, Ajello M, Allafort A, Antolini E, Atwood W B, et al. The second catalogue of active galactic nuclei detected by the Fermi large area telescope. ApJ. 2011; 743:171-208

Ackermann M, Ajello M, Atwood W B, Baldri E.et al. The third Catalog of Ative Galactic Nuclei Detected by the Fermi Large Area Telescope. Astrophys. J., 2015; 810, 14-47

AjelloM, AngioniR,Axelsson, M. Ballet, J. The Fourth Catalog of Active Galactic Nuclei Detected by the Fermi Large Area Telescope, *Astrophysical Journal*,2020; 892, 105

Alhassan, J.A., Ubachukwu, A. A, Odo F. C., Onuchukwu, C C. Relativistic beaming effects and structural asymmetries in highly asymmetric double radio sources. Rev MexAA. 2019; 55: 151-159

Alhassan J A, Ubachukwu, A A, and Odo F C. On the absence of core Luminosity-core Dominance Parameter ($P_c$-R) Correlation in Radio Galaxies and BL Lacs. J. Astrophy. & Astro., 2013; 34, 61-67

Alonso, M. F. (2019). Redshift of the Blazar KUV 00311-1938: Modeling the EBL Absorption. arXiv:1909.03960v1 [astro-ph.HE]. 2019

Athreya R M, Kapahi V K. The redshift dependence of spectral index in powerful radio galaxies. Jour. of Astrophys. & Astron.1998; 19: 63-77

Behera, B., and Wagner, S. J. (2009). Spectral trends in the Very High Energy blazar sample due to EBL absorption. arXiv:0901.1001v1 [astro-ph.HE], 2009

Blandford, R. D. and Rees, M. J, in Pittsburgh Conference on BL Lac Objects, ed. A. M. Wolfe (Pittsburgh: University of Pittsburgh), 328 1978

Böttcher M, and Chiang J. X-Ray Spectral Variability Signatures of Flares in BL Lacertae ObjectsAstrophys. J.,2002; 581, 127

Boula, S., Kazanas, D and Mastichiadis A. Accretion Disc MHD Winds and Blazar Classification MNRAS, 2019; 482, L80

Cellone S A, Romero G E, Araudo A T. Extremely violent optical micro-variability in blazars: fact or fiction? MNRAS. 2007; 374: 357- 364

Chen Y Y, Zhang X, Xiong D R, Wang S J,and Yu, X. L. The beaming effect and $\gamma$-ray Emission for Fermi blazars. Res. Astron. Astrophys. 2016; 16, 13

Cheng K S, Fan J H and Zhang L, Basic Properties of Gamma-ray Loud Blazars *A&A*, 1999; 352, 32

Dondi L and Ghisellini G. *γ*-ray-loud blazars and Beaming. MNRAS, 1995; 273, 583-595

Fan, J.H. The Basic Parameters of *γ*-ray-Loud Blazars. A&A, 2005; 436, 799

Fan X L, Bai J M and Mao, J R. What Determines the Observational Differences of Blazars? Res. Astron. Astrophys., 2016; 16,173

Finke J D. Compton Dominance and the Blazar sequence. Astrophysical J. 2013; 763, 134

Fossati G, Maraschi L, Celotti A, Comastri A and Ghisellini G. A Unifying View of the Spectral Energy Distributions of Blazars. MNRAS. 1998; 299, 433 – 448.

Ghisellini, G, Righi, C., Costamante, L., Tavecchio, F. The Fermi blazar sequence. Mon. Not. R. Astron. Soc. 2017; 234, 23-34

Ghisellini G, Tavecchio F, Foschini L and Ghirlanda G, The Transition Between BL Lac Objects and Flat Spectrum Radio Quasars.MNRAS, 2011; 414, 2674

Ghisellini G, Celotti, A, Fossati, G, Maraschi, L and Comastri, A. Theoretical Unifying Scheme for Gamma-ray Bright Blazars. MNRAS, 1998; 301, 451-468

Ghisellini G, Padovani P, Celotti A. and Maraschi, L. Relativistic Bulk Motion in Active Galactic Nuclei. Astrophysical J. 1993; 407, 65

Giommi P, Padovani, P., Polenta, G, Turriziani S, D'Elia V. et al. A Simplified View of Blazars: Clearing the Fog around Long-Standing Selection Effects. MNRAS. 2012; 420, 2899–2911

Giommi, P., Padovani, P., Polenta, G. A simplified view of blazars: the γ-ray case. MNRAS. 2013; 431, 1914

Iyida, E. U. Odo, F. C., Chukwude, A. E., Ubachukwu, A. A. Gamma-ray luminosity and spectral energy distributions of selected Fermi blazars. Publ. of the Astro. Soc. of Nig., 2019; 4, 74–81

Iyida, E.U. Odo, F.C., Chukwude, A.E. and Ubachukwu, A.A. Spectral Properties of *Fermi* Blazars and their Unification Schemes. 2020: *Open. Astro*. 29: 168-178

Iyida, E.U., Odo F.C. & Chukwude, A.E. 2021. Radio Core-Dominance of *Fermi*-Blazars: Implication for Blazar Unification. *Astrophys & Space Sci*. 366: 40

Kang S J, Fan, J H, Mao W, et al. Evaluating the Optical Classification of *Fermi* BCUs Using Machine Learning. Astrophysical J. 2019; 872, 189

Ledden JE, Odell SL. (1985). The radio-optical-x-ray spectral flux distributions of blazars. ApJ. 298: 630

Lind KR, Blandford R D. Semi-dynamical models of radio jets - relativistic beaming and source counts. ApJ. 1985; 295: 358

Mao P, Urry C M, Massaro, F and Paggi, A. A Comprehensive Statistical Description of Radio-through-*γ*-ray Spectral Energy Distributions of all Known Blazars. ApJS, 2016; 224, 26

Mastichiadis A, Kirk, J G. Variability in the synchrotron self-Compton model of blazar emission. A&A, 1997; 320, 19

Meyer E T, Fossati G, Georganopoulos, M, Lister M L. From the Blazar Sequence to the Blazar Envelope: Revisiting the Relativistic Jet Dichotomy in Radio-Loud Active Galactic Nuclei. Astrophys. J, 2011; 740, 98

Nalewajko K and Gupta M. The Sequence of Compton Dominance in Blazars based on Data from WISE and *Fermi*-LAT A&A, 2017; 606, A44

Odo, F. C., Ubachukwu, A. A. and Chukwude, A. E., Relativistic Beaming and Orientation Effects in BL Lac objects, *J. Astrophys. Astron*. 2012;**33**, 279

Odo F. C., Chukwude, A E, Ubachukwu, A A. Relativistic Beaming Effects in BL Lacertae Objects: Evidence for RBL/XBL Dichotomy. Astrophys. Space Sci., 2017; 362, 23

Odo F.C., Chukwude A.E., Ubachukwu A.A. Luminosity-redshift (*L- z*) Relation and the Blazar Sequence for Low Power Blazars. Astrophys. Space Sci., 2014; 349, 939

Odo F C and Aroh, B E. On the Unified View of Gamma-ray Energy Distribution of BL Lac Objects and Flat Spectrum Radio Quasars. J. Astrophy. & Astro., 2020; 41, 9

Ostriker J P and Vietri M, Are Some BL Lac Objects Artefacts of Gravitational Lensing? Nature, 1985; 318, 446

Padovani, P. Is there a Relationship between BL Lacertae Objects and Flat-Spectrum Radio Quasars? MNRAS, 1992; 257, 404

Padovani P.Urry C.M., Blazar Demographics and Physics, *ASP Conf Series*, 2001; 227

Padovani P. The blazar sequence: validity and predictions. Astrophy. & Space Scie, 2007; 309: 63-71

Palladino A, Rodrigues X, Gao S, Winter W. Interpretation of the Diffuse Astrophysical Neutrino Flux in Terms of the Blazar Sequence. Astrophys. J. 2019; 871, 41

Pavlidou, V., Richards, J.L. Max-Moerbeck, W., King, O.G., Pearson, T.J. Readhead, A.C.S. Assessing the Significance of apparent correlations between radio and gamma-ray blazar fluxes. ApJ, 2012; 751, 149-161

Pei ZY, Fan JH, Bastieri D, Sawangwit U, Yang J H. The relationship between the radio core-dominance parameter and spectral index in different classes of extragalactic radio sources (II). Res. Astron. Astrophys. 2019; 19: 70-86

Press, W.H., Teukolsky, S.A., Vetterling, W.T., Flannery, B.P., 1994, Numerical Recipes in Fortran, The art of Scientific Computing, Second Edition, Cambridge University Press

Ouyang, Z, Xiao, H., Zheng, Y., Xu, H., Fan, J. (2021). The spectral index study for Fermi blazars. *Astrophys Space Sci.* 366:12

Sambruna R M, Maraschi L, Urry, C M. On the Spectral Energy Distributions of Blazars. Astrophysical Journal. 1996; 463, 444

Savolainen, T., Homan, D.C., Hovatta, T., Kadler, M. et al.Relativistic Beaming and Gamma-ray Brightness of Blazars. Astro & Astro. 2010; 512, A24

Ubachukwu, A A, Chukwude A E, On the relativistic beaming and Orientation effects in core-dominated quasars. J. Astrophys&Astr., 2002; 23, 235-242

Urry C M and Padovani P. Unified Schemes for Radio-Loud Active Galactic Nuclei. Publ. Astron. Soc. Pacific. 1995; 107, 803

Vagnetti F, Giallongo E and Cavaliere A. BL Lacertae Objects and Radio-Loud Quasars Within an Evolutionary Unified Scheme. Astrophysical J. 1991; 368, 366 – 372

Xie G Z, Dai B Z, Mei D C, Fan J H. The Doppler Effect and Spectral Energy Distribution of Blazars. Chin. J. Astron. Astrophys. 2001; 1, 213-220

Xie G Z, Liu F K, Liu, The Beaming Model and Hubble Diagram of BL Lacertae objects, A&A, 1991; 249, 65

Yan D, Zeng H and Zhang L. The Physical Properties of Fermi BL Lac Objects Jets. MNRAS, 2014; 439, 2933

Zhang, J., Liang, E-W., Zhang, S.-N., & Bai, J. M. Radiation Mechanisms and Physical Properties of GeV-TeV BL Lac Objects. ApJ, 2012; 752, 157

Zheng Y G, Yang C Y, Zhang L and Wang J. C. Discerning the Gamma-Ray-Emitting Region in The Flat Spectrum Radio QuasarsApJS, 2017; 228,1